\documentclass[aps,preprint,showpacs,amsmath,amssymb]{revtex4}
\usepackage{graphicx}
\usepackage{amssymb}
\usepackage{bm}
\usepackage{subfigure}  

\begin{document}
\setcounter{page}{1}

\title{Effect of photonic band gap on entanglement dynamics of qubits}

\author{Jing-Nuo \surname{Wu$^{1}$}}
\email {jingnuowu@gmail.com}
\thanks{FAX:+886-2-28610577}
\author{Wen-Feng \surname{Hsieh$^{2}$}}
\author{Szu-Cheng \surname{Cheng$^{1}$}}
\email {sccheng@faculty.pccu.edu.tw}
\thanks{FAX:+886-2-28610577}

\affiliation{$^{1}$Department of Physics, Chinese Culture University, Taipei, Taiwan, R. O. C.}
\affiliation{$^{2}$Department of Photonics and Institute of Electro-Optical Engineering, National Chiao Tung University, Hsinchu, Taiwan, R. O. C.}
\date[]{Received \today }

\begin{abstract}
We study how the environment of photonic band gap (PBG) materials affects entanglement dynamics of qubits. Entanglement between the single qubit and the PBG environment is investigated through the von Neumann entropy while that for two initially entangled qubits in this PBG reservoir is through concurrence. Dynamics of these measurements are solved in use of the fractional calculus which has been shown appropriate for the systems with non-Markovian dynamics. Entropy dynamics of the single qubit system reveals that the coupling with the PBG reservoir prevents decoherence of the qubit through the steady entropy with non-zero value. The effect of PBG reservoir on the concurrence of the two-qubit system leads to the long-time entanglement preservation. The concurrence dynamics shows that unphysical entanglement trapping does not exist in the system with the qubit frequency lying outside the PBG region. Long-time memory effect of the PBG reservoir occurs only for the qubit frequency in the PBG region. Entanglement mechanisms resulting from this long-time memory effect are discussed.    
   
\end{abstract}
\maketitle

\section{INTRODUCTION}
Photonic band gap (PBG) materials, in which no traveling electromagnetic (EM) modes are allowed in the PBG of their band structures, enrich the content of quantum optics and quantum information processing through its non-Markovian dynamics. Periodic structures constructing these PBG materials posses the controllable features by the lattice parameters.  The development of these PBG materials has attracted the interest over the past twenty years starting from Yablonovitch's seminar paper \cite{Yablonovitch87}, where the periodic structures with dielectric variation of photonic crystals (PhCs) were proposed to be a means to inhibit spontaneous emission (SE). Shortly thereafter, John \textit{el} \textit{al}. found that the PhC mediates a coherent interaction between the emitter and the previously emitted radiation \cite{John90}. The memory effect in this coherent interaction leads to non-Markovian emission dynamics such as non-exponential decay and Rabi-like oscillation in the spectral vicinity of a photonic band edge \cite{John94}. This non-Markovian emission dynamics enriches the content of quantum optics and quantum information processing in the PBG materials. Recently, tunable artificial crystals consisting of Josephson junction arrays have been designed as a PBG structure with tunable frequency range through the applied external flux \cite{Savel05, Nori08, Hutter11}. This PBG structure of superconducting circuits with more adjustable parameters than PhCs might ultimately form the primitive blocks of quantum computers. 

When two initially entangled qubits are placed in the PBG materials, it has been reported that the controllable SE of the individual qubit can be used to manipulate the entanglement of the two qubits. For example, Bellomo \textit{et} \textit{al.} \cite{Bellomo08} suggested high value of entanglement trapping between two qubits can be achieved by embedding in an isotropic PBG environment. The direct link between the entanglement dynamics and the controllable SE dynamics of the single qubit is the basis of achieving entanglement preservation. Besides, Amri \textit{et} \textit{al.} \cite{Amri09} in exploring the entanglement dynamics of two qubits embedded in isotropic and anisotropic PhCs found that the initially entangled qubits will become disentangled as the excited qubits spontaneously emit the photons into the environment. Whenever the individual qubit forms the bound state with the emitted photon, the initial entanglement between the two qubits can be partially recovered. Deep understanding about what the entanglement mechanism is between the qubits by the PBG reservoir through the non-Markovian dynamics of the qubits is now indispensable due to the possibilities for realizing strong-coupling conditions in the exploration systems of quantum information processing \cite{Nataf10, Tian10}.  

Here we consider the system of qubits placed in a PBG material such as a PhC or tunable artificial crystal with directional-dependent dispersion relation near the photonic band edge. This anisotropic band structure of the PBG reservoir can be described by the effective-mass approximation \cite{SJohn87, Wu2010} and leads to the photon density of state (DOS) of the reservoir  proportional to $\sqrt{\omega-\omega_{c}}\Theta(\omega-\omega_{c})$, where $\omega_{c}$ is the band edge frequency and $\Theta(\omega-\omega_{c})$ is the Heaviside step function. Since the traveling EM modes are forbidden in the PBG region, the emitted photon from the qubit can't travel through the PBG material and thus is localized around the qubit to form a bound state. This bound state carries the memory of the PBG reservoir on the excited-state qubit. This memory effect controls the dynamics of the qubit through the delay Green function or memory kernel. The fractional calculus \cite{AAStanislavsky04} has been shown an appropriate mathematical method for solving these physical systems with memory effect. Using the fractional calculus to investigate the non-Markovian SE dynamics of the PhCs we obtained the results consistent with the experimental observation \cite{SCCheng09, Wu2010}.

Here we use the fractional calculus to investigate how the entanglement dynamics of the qubits are affected by the memory effect of the PBG reservoir. The systems we consider include one single qubit and two entangled qubits surrounded by the PBG materials, respectively. The entanglement between the single qubit and the PBG reservoir is investigated through the qubit's excited-state probability and von Neumann entropy while the entanglement between the two qubits is through the concurrence. The probability dynamics of the single qubit exhibits non-Markovian and Markovian behavior as the qubit frequency lies inside and outside the PBG region, respectively. The memory effect of the PBG reservoir on the excited-state qubit leads to the non-Markovian dynamics of the qubit through the formation of photon bound state. This memory effect leads to the entanglement preservation between the single qubit and the PBG reservoir through the steady entropy with non-zero value. In the two-qubit system, we find that unphysical entanglement trapping does not exist in the system with qubit frequency lying outside the PBG region. These new results of the PBG on the entanglement dynamics are different from those of previous studies which predicted the existence of entanglement trapping in the system with qubit frequency outside the PBG region \cite{Bellomo08, Amri09}. The accuracy of these results is certified by the use of appropriate mathematical method of fractional calculus for non-Markovian dynamics and the reasonable results of physical phenomenon. Besides, we discuss the entanglement mechanisms for the two-qubit system with different prepared states. PBG reservoirs provide the qubits a better environment for preserving entanglement through interchanging the trapped photon. These accurate results about the effect of PBG reservoir on the qubit's dynamics can be applied directly to the experimental systems of tunable artificial crystals with PBG structure through the circuit quantum electrodynamics (circuit QED) setup. These artificial crystals, composed of periodic arrays of circuit elements of Josephson junction, have been designed to have a gap in the EM spectrum through the array parameters \cite{Wallraff04, Savel05, Nori08, Fink09, Neeley10, DiCarlo10, Hutter11}. As qubits are placed in the middle of these arrays of artificial crystals, the emitted photons from the qubits in the PBG environment is equivalent to that of qubits embedded inside a PhC.

The paper is organized as follows. In Sec. II, starting from the quantum theory of one single qubit in an anisotropic PhC, we express the kinetic equation of this interaction system as a fractional Langevin equation in use of the fractional calculus. By solving the fractional Langevin equation analytically, we express the time evolution of the qubit's quantum state as a reduced density matrix. Dynamics of probability and von Neumann entropy associated with the matrix elements are discussed.  In Sec. III, we construct the reduced density matrix of the two-qubit system through the single-qubit matrix and the two-qubit basis. The entanglement dynamics of the two-qubit system is discussed on the basis of the concurrences of two initially entangled states. The entanglement mechanisms for the non-Markovian systems with different prepared states are revealed in this section. Finally, we summarize our results in Sec. IV.

\section{Probability and von Neumann entropy of one single qubit}
As a single qubit is surrounded by an anisotropic PBG material with cut-off photonic DOS shown in Fig. 1, the total Hamiltonian of this interaction system can be expressed as
\begin{equation}
H=\frac{1}{2}\hbar\omega_{10}(\left|1\right\rangle\left\langle 1\right|-\left|0\right\rangle\left\langle 0\right|)+\sum_{\vec{k}}{\hbar\omega_{\vec{k}}a_{\vec{k}}^{+}a_{\vec{k}}}+i\hbar\sum_{\vec{k}}{g_{\vec{k}}(\left|0\right\rangle\left\langle 1\right|\otimes a_{\vec{k}}^{+}+{h.c.})},
\end{equation}
where $|{1}\rangle$ and $|{0}\rangle$ stand for the excited and ground states of the qubit with transition frequency $\omega_{10}$ nearly resonant with the frequency range of PBG; $a_{\vec{k}}$ and $a_{\vec{k}}^{+}$ are the annihilation and creation operators of photon with wavevector $\vec{k}$ and corresponding frequency $\omega_{\vec{k}}$ in the reservoir; coupling strength between the qubit and the $\vec{k}$ photon is specified by $g_{\vec{k}}=\frac{\omega_{10}d_{10}}{\hbar}[{\frac{\hbar}{2\epsilon_{0}\omega_{\vec{k}}V}]}^{\frac{1}{2}}\hat{e}_{\vec{k}}\cdot\hat{u}_d$ , which is assumed to be independent of the qubit's position with fixed qubit's dipole moment $\vec{d}_{10}=d_{10}\hat{u}_d$. Here the symbols $V$, $\hat{e}_{\vec{k}}$ and $\epsilon_{0}$ stand for the sample volume, polarization unit vector of the $\vec{k}$ photon and the dielectric constant, respectively.

If we use the coordinate ($\theta,\phi$) on the Bloch sphere to parameterize the state of the qubit, the initial state can be written as
\begin{equation}
|\psi(0)\rangle=\left[e^{i\phi_{0}}cos(\frac{\theta_{0}}{2})|1\rangle+sin(\frac{\theta_{0}}{2})|0\rangle\right]\otimes \left|\textbf{0}_{\vec{k}}\right\rangle+\sum_{\vec{k}}{C_{\vec{k}}(0)|0\rangle\otimes \left|\textbf{1}_{\vec{k}}\right\rangle}
\end{equation}
with the initial coordinate ($\theta_{0},\phi_{0}$)=$(0,\phi)$ for the excited state and $(\pi,\phi)$ for the ground state. Here the photon vacuum state and one-photon state are expressed as $\left|\textbf{0}_{\vec{k}}\right\rangle$ and $\left|\textbf{1}_{\vec{k}}\right\rangle$, respectively. We assume $C_{\vec{k}}(0)=0$ for no initial correlation between the qubit and PBG reservoir. This system evolves with time through
\begin{equation}
|\psi(t)\rangle=\left[u_{p}(t)e^{i\phi_{0}}cos(\frac{\theta_{0}}{2})|1\rangle+u_{d}(t)sin(\frac{\theta_{0}}{2})|0\rangle\right]\otimes \left|\textbf{0}_{\vec{k}}\right\rangle+\sum_{\vec{k}}{C_{\vec{k}}(t)|0\rangle\otimes \left|\textbf{1}_{\vec{k}}\right\rangle}
\end{equation}
with initial condition $u_{p}(0)=1$, $u_{d}(0)=1$ and $C_{\vec{k}}(0)=0$. Here $u_{p}(t)$ and $u_{d}(t)$ stand for the excited-state and ground-state probability amplitudes of the qubit with the photon vacuum state while $C_{\vec{k}}(t)$ for the qubit in its ground state with one photon in the reservoir.

Substituting Eq. (3) into the time-dependent Schr$\ddot{o}$dinger equation then taking inner product with $\left\langle\psi(t)\right|$, we obtain the coupled amplitude equations as
\begin{equation}
\frac{d}{dt}u_{p}(t)=-\frac{1}{cos(\theta_{0}/2)}\sum_{\vec{k}}{g_{\vec{k}}C_{\vec{k}}(t)e^{-i\Omega_{\vec{k}}t}},
\end{equation}
\begin{equation}
\frac{d}{dt}C_{\vec{k}}(t)=g_{\vec{k}}u_{p}(t)cos(\theta_{0}/2)e^{i\Omega_{\vec{k}}t},
\end{equation}
and
\begin{equation}
\frac{d}{dt}u_{d}(t)=0
\end{equation}
with the detuning frequency $\Omega_{\vec{k}}=\omega_{\vec{k}}-\omega_{10}$.

The last equation of motion yields $u_{d}(t)=u_{d}(0)=1$ meaning that the ground-state probability amplitude will not evolve with time. Integrating Eq. (5) and substituting into Eq. (4), we have
\begin{equation}
\frac{d}{dt}u_{p}(t)=-\int_{0}^{t}{G(t-\tau)u_{p}(\tau)d\tau}
\end{equation}
with the memory kernel
$G(t-\tau)=\sum_{\vec{k}}{g_{\vec{k}}^{2}e^{-i\Omega_{\vec{k}}(t-\tau)}}$. This equation reveals that the future state of the qubit is related to the memory of the reservoir in its previous state through the memory kernel. As the qubit is put in free space, the memory kernel has the form of a Dirac delta function $G(t-\tau)\propto\delta(t-\tau)$ corresponding to continuous photon DOS. In this case, the reservoir responses only at an instant time $\tau=t$ which leads to the excited amplitude of the qubit decaying exponentially with time. This Markovian result in free space manifests that the qubit loses all memory of its past and decays quickly to its ground state.

For the anisotropic PBG reservoir discussed here, the memory kernel manifests its memory effect within the entire interval (0,t) through $G(t-\tau)=\frac{\beta^{1/2}/f^{3/2}}{\sqrt{\pi}(t-\tau)^{3/2}}e^{-i[3\pi/4-\delta(t-\tau)]}$  with the coupling constant $\beta^{1/2}=(\omega^{2}_{10}d^{2}_{10}\sqrt{\omega_{c}})/(16\pi\epsilon_{0}\hbar c^{3})$ and the detuning $\delta=\omega_{10}-\omega_{c}$ of the qubit frequency $\omega_{10}$ from the band edge frequency $\omega_{c}$ \cite{Wu2010}. This memory kernel is derived from the anisotropic dispersion relation of the PBG material. Near the band edge frequency $\omega_{c}$, the anisotropic dispersion relation has a vector form and can be expressed by the effective-mass approximation as \cite{SJohn87} $\omega_{\vec{k}}\approx\omega_{c}+A\left(\vec{k}-\vec{k}_{c}\right)^{2}$ with the curvature $A\cong f\omega_{c}/k^{2}_{c}=fc^{2}/\omega_{c}$ signifying its directional-dependent values through the scaling factor $f$. This dispersion relation leads to the memory kernel expressed by the cut-off photon DOS   $D(\omega)=\frac{1}{4\pi^{2}}\sqrt{\frac{\omega-\omega_{c}}{A^{3}}}\Theta(\omega-\omega_{c})$ as $G(t-\tau)=\frac{\omega^{2}_{10}d^{2}_{10}}{4\epsilon_{0}\hbar}\int^{\infty}_{0}d\omega\frac{D(\omega)}{\omega}e^{-i(\omega-\omega_{10})(t-\tau)}$. Substituting this memory kernel for the anisotropic PBG reservoir into Eq. (7), we obtain
\begin{equation}
\frac{d}{dt}u_{p}(t)=-\frac{\beta^{1/2}e^{i3\pi/4}}{\sqrt{\pi}f^{3/2}}\int^{t}_{0}\frac{u_{p}(\tau)e^{i\delta(t-\tau)}}{(t-\tau)^{3/2}}d\tau.
\end{equation}
This equation can be further simplified by making the transformation $u_{p}(t)=e^{i\delta t}U_{p}(t)$ to give
\begin{equation}
\frac{d}{dt}U_{p}(t)+i\delta U_{p}(t)=\frac{\beta^{1/2}e^{i\pi/4}}{\sqrt{\pi}f^{3/2}}\int^{t}_{0}\frac{U_{p}(\tau)}{(t-\tau)^{3/2}}d\tau.
\end{equation}
Conventionally, this integro-differential equation is solved through Laplace transform which leads to the fractal phenomenon of the system for the memory kernel in Laplace image \cite{RRNigmatullin92}. This fractal phenomenon would result in the stochastic nature of the dynamical behavior of the system which appears as the non-Markovian dynamics. This non-Markovian dynamics can be solved analytically by the fractional calculus.

Here we solve the non-Markovian dynamics of the single qubit system by the fractional calculus. The detailed derivation can be found in Appendix A. We obtain the analytical solution of Eq. (9) as
\begin{equation}
U_{p}(t)=\frac{1}{2e^{i\pi/4}\sqrt{\beta/f^{3}-\delta}}\times\left[Y^{2}_{1}E_{t}(1/2,Y^{2}_{1})-Y^{2}_{2}E_{t}(1/2,Y^{2}_{2})+Y_{1}e^{Y^{2}_{1}t}-Y_{2}e^{Y^{2}_{2}t}\right],
\end{equation}
for $\beta/f^{3}\neq\delta$; and
\begin{equation}
U_{p}(t)=-2\frac{\beta^{3/2}e^{i3\pi/4}}{f^{9/2}}tE_{t}(\frac{1}{2},i\beta/f^{3})-\frac{\beta^{1/2}e^{i\pi/4}}{f^{3/2}}E_{t}(\frac{1}{2},i\beta/f^{3})
+(1+\frac{2it\beta}{f^{3}})e^{i\beta t/f^{3}}-2\frac{\beta^{1/2}e^{i\pi/4}}{f^{3/2}\sqrt{\pi}}t^{1/2}
\end{equation}
for $\beta/f^{3}=\delta$. 
The time evolution of the system is thus analytically expressed as
\begin{equation}
|\psi(t)\rangle=\left[u_{p}(t)e^{i\phi_{0}}cos(\frac{\theta_{0}}{2})|1\rangle+u_{d}(t)sin(\frac{\theta_{0}}{2})|0\rangle\right]\otimes \left|\textbf{0}_{\vec{k}}\right\rangle+\sum_{\vec{k}}{C_{\vec{k}}(t)|0\rangle\otimes \left|\textbf{1}_{\vec{k}}\right\rangle}
\end{equation}
with $u_{p}(t)=e^{i\delta t}U_{p}(t)$, $u_{d}(t)=u_{d}(0)=1$ and $\sum_{\vec{k}}\left|C_{\vec{k}}(t)\right|^{2}=1-\left[u_{p}(t)cos(\frac{\theta_{0}}{2})\right]^{2}-\left[sin(\frac{\theta_{0}}{2})\right]^2$. The reduced density matrix of a single qubit in the anisotropic PBG reservoir can also be expressed analytically as
\begin{equation}
\hat{\rho}^{s}(t)=\bordermatrix{ \cr  &  \cr
  & \left|u_{p}(t)\right|^{2}cos^{2}(\frac{\theta_{0}}{2}) &  \frac{1}{2}u^{*}_{p}(t)e^{-i\phi_{0}}sin(\theta_{0}) \cr   &  \frac{1}{2}u_{p}(t)e^{i\phi_{0}}sin(\theta_{0})  & 1-\left|u_{p}(t)\right|^{2}cos^{2}(\frac{\theta_{0}}{2}) \cr}\equiv\bordermatrix{ \cr  &  \cr
  & \rho_{11}(t) &  \rho_{10}(t) \cr   & \rho_{01}(t)  & \rho_{00}(t) \cr}
\end{equation}
with the initial one $\hat{\rho}^{s}(0)=\bordermatrix{ \cr  &  \cr
  & cos^{2}(\frac{\theta_{0}}{2}) &  \frac{1}{2}e^{i\phi_{0}}sin(\theta_{0}) \cr   & \frac{1}{2}e^{-i\phi_{0}}sin(\theta_{0})  & sin^{2}(\frac{\theta_{0}}{2}) \cr}$.
  
We show the effect of PBG reservoir on the excited-state probability and von Neumann entropy of the single qubit in Fig. 2. For the qubit frequency $\omega_{10}$ lies above the edge frequency $\omega_{c}$ or the positive detuning case [$\delta/\beta=(\omega_{10}-\omega_{c})/\beta>0$], the excited-state probability exhibits Markovian behavior with exponential decay. The initially excited qubit loses almost its energy into the PBG reservoir and decays to the ground state. During this decaying process, the qubit loses all memory of its past through emitting a photon. For the negative detuning case with the qubit frequency lying inside the PBG region [$\delta/\beta=(\omega_{10}-\omega_{c})/\beta<0$], the system manifests decay and oscillatory behavior before reaching a non-zero steady-state value. This non-Markovian dynamics reveals that the qubit loses partial of its energy in the very beginning period of time and then preserves the remaining energy in the end. In this case, the emitted photon from the qubit 'sees' no allowed DOS in the PBG region so that it is localized around the qubit through the reflecting and re-absorbed processes and forms the bound state with the qubit. The formation of the bound state preserves the remaining energy of the qubit and the memory of the PBG reservoir on the excited qubit. This memory effect leads to the distinct dynamical behavior of the non-Markovian system from the Markovian one. When a quantum operation with multiple stages is performed on the qubit, the system with memory effect will experience non-Markovian processes of operation with the pre-stage operation consecutively affecting the following one. On the other hand, the qubit in the Markovian system without memory effect will experience stochastic processes of operation with each stage operation independent of the other.     

In Fig. 2(b), we show the dynamics of von Neumann entropy in terms of the eigenvalues of the reduced density matrix in Eq. (13).  This quantum entropy is defined as $S(t)=-Tr\left[\hat{\rho}^{s}(t) log\hat{\rho}^{s}(t)\right]=-\sum_{i}\lambda_{i}log\lambda_{i}$, where $\lambda_{i}$ are the eigenvalues of the density matrix $\hat{\rho}^{s}(t)$. The entropy measures the correlation between the reservoir and the qubit through quantifying the amount of the information we will gain after we measure a quantum state. The larger the entropy value is, the stronger correlation exists. For the density matrix in Eq. (13), the eigenvalues can be obtained easily as $\lambda_{\pm}=\frac{1}{2}\left\{1\pm\sqrt{1-4cos^{4}\left(\frac{\theta_{0}}{2}\right)\left[\left|u_{p}(t)\right|^{2}-\left|u_{p}(t)\right|^{4}\right]}\right\}$. We plot the entropy dynamics in Fig. 2(b) based on these eigenvalues. This quantum entropy evolves with time because the correlation with the PBG reservoir leads to the amount of information varying with time. For the initial pure state of the excited qubit, the entropy exhibits its minimal value of zero at $t=0$ and reaches its maximal value $log 2=0.693$ at the very beginning of time. After a period of time of the decay timescale, the entropy becomes steady with non-zero value in the non-Markovian system ($\delta/\beta<0$) and zero in the Markovian system ($\delta/\beta=2$). These results show that the initial pure system becomes maximally mixed in the very beginning period of time because of correlation with the reservoir. As the qubit equilibrates with the PBG reservoir, the system becomes less mixed. Contrarily, in the Markovian system, the qubit decays to its another pure state (ground state), as it quickly equilibrates with the reservoir. In this case, the stored information in the qubit is lost because of the correlation with the Markovian reservoir \cite{Nielsen00}. However, in the non-Markovian system, the strong correlation with the PBG reservoir leads to the preservation of the mixed state. The larger non-zero steady entropy value of the non-Markovian system has, the greater amount of information we will gain upon measuring this mixed state. Since the system with qubit frequency deeper inside PBG region has the larger non-zero steady value of entropy, we can use the absolute value of qubit's detuning frequency $\left|\delta\right|/\beta$ to characterize the degree of the non-Markovian effect from PBG reservoir. Correlation with the PBG reservoir prevents losing stored information from the single qubit through the large value of the steady entropy.

\section{Entanglement of two qubits}
In the following we consider a system consisting of two independent parts, each with a qubit coupled to the PBG reservoir with anisotropic one-band model. The two qubits $A$ and $B$ with the same transition frequency $\omega_{10}$ are assumed to be identical and initially entangled. The distance between the two qubits is larger than the spatial correlation length between the individual qubit and PBG reservoir. For each part of the subsystem with one qubit and the corresponding PBG reservoir, the reduced density matrix for the single qubit can be described by Eq. (13) with the superscript $s=A,B$. For the two-qubit system, the reduced density matrix $\hat{\rho}(t)$ can be constructed through the single-qubit matrix $\hat{\rho}^{s}(t)$ and the two-qubit basis ${\left|11\right\rangle\equiv\left|1\right\rangle, \left|10\right\rangle\equiv\left|2\right\rangle, \left|01\right\rangle\equiv\left|3\right\rangle, \left|00\right\rangle\equiv\left|4\right\rangle}$ \cite{Bellomo07, Bellomo08}. This reduced density matrix for the two-qubit system is a Hermitian matrix with its elements $\rho_{ij}(t)$, $i,j=1~4$, depending only on the initially entangled condition and the excited-state amplitude $u_{p}(t)$ of the single qubit. For the initially entangled states $ \left|\Phi\right\rangle=\alpha\left|01\right\rangle+\gamma\left|10\right\rangle=\alpha\left|3\right\rangle+\gamma\left|2\right\rangle$ and $\left|\Psi\right\rangle=\alpha\left|00\right\rangle+\gamma\left|11\right\rangle=\alpha\left|4\right\rangle+\gamma\left|1\right\rangle$ with $\alpha$ being real, $\gamma=\left|\gamma\right|e^{i\varphi}$ and $\alpha^{2}+\left|\gamma\right|^{2}=1$, the matrix elements are $\rho_{11}(t)=\rho_{11}(0)\left|u_{p}(t)\right|^{4}$, $\rho_{22}(t)=\rho_{11}(0)\left|u_{p}(t)\right|^{2}\left[1-\left|u_{p}(t)\right|^{2}\right]+\rho_{22}(0)\left|u_{p}(t)\right|^{2}$, $\rho_{33}(t)=\rho_{11}(0)\left|u_{p}(t)\right|^{2}\left[1-\left|u_{p}(t)\right|^{2}\right]+\rho_{33}(0)\left|u_{p}(t)\right|^{2}$, $\rho_{44}(t)=1-\left[\rho_{11}(t)+\rho_{22}(t)+\rho_{33}(t)\right]$, $\rho_{14}(t)=\rho_{14}(0)\left|u_{p}(t)\right|^{2}$ and $\rho_{23}(t)=\rho_{23}(0)\left|u_{p}(t)\right|^{2}$. Here the initial elements are $\rho_{ij}(0)=\left(\left|\Phi\right\rangle\left\langle \Phi\right|\right)_{ij}$ and $\left(\left|\Psi\right\rangle\left\langle \Psi\right|\right)_{ij}$ for the prepared $\left|\Phi\right\rangle$ and $\left|\Psi\right\rangle$ states.

In order to investigate how the PBG reservoir affects the entanglement of the two qubits, we calculate the concurrence of the system which provides the direct measurement of the entanglement \cite{Enk09}. For the initially entangled states $\left|\Phi\right\rangle$ and $\left|\Psi\right\rangle$, the concurrences are given by \cite{Wootters98, Bellomo08}
\begin{equation}
C_{\Phi}(t)= max. \left\{ 0, 2\left|\rho_{23}(t)\right|-2\sqrt{\rho_{11}(t)\rho_{44}(t)} \right\}
\end{equation}
and
\begin{equation}
C_{\Psi}(t)= max. \left\{ 0, 2\left|\rho_{14}(t)\right|-2\sqrt{\rho_{22}(t)\rho_{33}(t)} \right\}.
\end{equation}
Substituting the above matrix elements into the concurrences, we arrive at 
\begin{equation}
C_{\Phi}(t)= max. \left\{ 0, 2\alpha\sqrt{1-\alpha^{2}}\left|u_{p}(t)\right|^{2} \right\}
\end{equation}
and
\begin{equation}
C_{\Psi}(t)= max. \left\{ 0, 2\sqrt{1-\alpha^{2}}\left|u_{p}(t)\right|^{2}\left\{\alpha-\sqrt{1-\alpha^{2}}\left[1-\left|u_{p}(t)\right|^{2}\right]\right\} \right\}. 
\end{equation}
These expressions of the concurrences have the neat forms of $C_{\Phi}(t)=\left|u_{p}(t)\right|^{2}$ and $C_{\Psi}(t)=\left|u_{p}(t)\right|^{4}$ if the two qubits are maximally entangled through the Bell states $\left|\Phi\right\rangle_{B}=(\left|01\right\rangle\pm\left|10\right\rangle)/\sqrt{2}$ and $\left|\Psi\right\rangle_{B}=(\left|00\right\rangle\pm\left|11\right\rangle)/\sqrt{2}$. 

In Fig. 3, we show the concurrence dynamics of the two qubits varying with the initial entangled degree $\alpha^{2}$ of the prepared states $\left|\Phi\right\rangle$ and $\left|\Psi\right\rangle$. For the prepared $\left|\Phi\right\rangle$ state, the concurrences with qubit frequency lying inside and outside PBG are respectively displayed in Fig. 3(a) (non-Markovian system, $\delta/\beta=-5$) and Fig. 3(b) (Markovian system, $\delta/\beta=2$) while those for the prepared $\left|\Psi\right\rangle$ state are in Fig. 3(c) (non-Markovian system, $\delta/\beta=-5$) and Fig. 3(d) (Markovian system, $\delta/\beta=2$). 
The non-Markovian system for the prepared $\left|\Phi\right\rangle$ state in Fig. 3(a) exhibits promising entanglement trapping for all initial entangled degree $\alpha\neq 0$. This entanglement trapping phenomenon is linked to the excited-state probability $P(t)=\left|u_{p}(t)\right|^{2}$ of the single qubit in steady bound state with non-zero value. This behavior induced by the memory effect of the PBG reservoir on the excited single qubit is evidently different from that of the Markovian system in Fig. 3(b), where $C_{\Phi}(t)$ decays exponentially and vanishes permanently. The backaction effect of the PBG reservoir on the non-Markovian system leads to the memory effect of the single qubit system and the entanglement trapping phenomenon in the two-qubit system. Besides the backaction effect of the PBG reservoir, the dependence of the entanglement dynamics on the initially entangled degree $\alpha^{2}$ is also clearly visible in Fig. 3(a). The stronger the initial entanglement is, the larger steady value of the concurrence reaches. It thus can be deduced from Fig. 3(a) that the maximally entangled $\left|\Phi\right\rangle$ state occurs at $\alpha^{2}=1/2$ which leads to the Bell states $\left|\Phi\right\rangle_{B}=(\left|10\right\rangle\pm\left|01\right\rangle)/\sqrt{2}$. Entanglement prefers to exist in the system with the two qubits sharing just one excitation \cite{SManiscalco08}. This fact of Bell state being maximally entangled $\left|\Phi\right\rangle$ state is also revealed in the Markovian system of Fig. 3(b). The initial entanglement of this Markovian system is followed by a period of complete disentanglement. The period of time it takes to reach the disentanglement depends on the degree of the initial entanglement. This time period is the longest as the initial entangled degree $\alpha^{2}$ equal to $1/2$. The fact that the system with two qubits sharing just one excitation exhibits maximally entangled reveals that the two qubits preserve their entanglement through interchanging the trapped photon in the PBG reservoir. Since the strength of this interchange is associated with the relative position of the two qubits, we believe the entanglement in the PBG system depends strongly on the position of the two qubits. This will be our future topic to discuss how the relative position of the qubits in the PBG material affects the entanglement of the system.  

Entanglement dynamics for the prepared $\left|\Psi\right\rangle$ state in Fig. 3(c) exhibits the similar behavior to those in Fig. 3(a) where entanglement trapping occurs for $\alpha^{2}\geq 1/3$. However, for $\alpha^{2}<1/3$, entanglement of the non-Markovian system vanishes permanently after a short period of time, similar to the Markovian case in Fig. 3(d) \cite{Yu06, Santos06}. For the prepared $\left|\Psi\right\rangle$ state in Figs. 3(c) and (d), the maximally entangled state occurs at $\alpha^{2}\cong 0.88$. The system is maximally entangled when the two qubits are almost in the same state coupled with two trapped photons. This fact reveals that the system preserves the entanglement through the dipole-dipole interaction (DDI) between the two qubits. The DDI has large strength when the dipoles of the two qubits have the same orientations. The stronger DDI between the two qubits, the better entanglement trapping the system exhibits.

When we further compare the non-Markovian systems in Fig. 3(a) and (c), we find that the dynamical difference between these two systems stems from the different entanglement mechanisms. PBG reservoir provides the qubits a better environment for preserving entanglement through interchanging the trapped photon than through DDI. This fact is more evident in Fig. 4(a) and (b) where we show the concurrence dynamics for the prepared Bell states  $\left|\Phi\right\rangle_{B}=(\left|10\right\rangle\pm\left|10\right\rangle)/\sqrt{2}$ and  $\left|\Psi\right\rangle_{B}=(\left|11\right\rangle\pm\left|00\right\rangle)/\sqrt{2}$. The steady values of the entanglement trapping for the prepared Bell state $\left|\Phi\right\rangle$ are larger than those for the $\left|\Psi\right\rangle$ state with the same detuning frequency $\delta/\beta$. This fact implies that the product of the probability of the states with only one excited qubit ($\left|10\right\rangle$, $\left|01\right\rangle$) is larger than that of the totally excited state ($\left|11\right\rangle$) and ground state ($\left|00\right\rangle$) of the two-qubit system. The larger product of probability of the states with only one excited qubit corresponds to the greater strength of interchanging the trapped photon than strength of DDI. The same result can also be observed in Fig. 4(c) where we show the steady values of the concurrences for the two prepared Bell states. For the same qubit's detuning frequency $\delta/\beta$, the system with prepared $\left|\Phi\right\rangle_{B}$ state exhibits larger steady values than those of the $\left|\Psi\right\rangle_{B}$ state which results from the greater strength of interchanging the trapped photon. 

When we consider the different detuning frequencies of the qubits in Fig. 4(c), the system displays the entanglement trapping only for the frequency lying within the PBG. That is, there does not exist entanglement trapping in the system with the qubit frequency lying outside PBG region. The strength of entanglement trapping can be indexed by the absolute value of detuning frequency $\left|\delta\right|/\beta$ inside the PBG region. The deeper the qubit frequency is detuned inside the PBG region, the stronger entanglement trapping yields.

\section{Conclusion}
We have investigated the effect of PBG on the dynamics of qubits surrounded by the PBG reservoir. In use of the fractional calculus, we solve the qubit's dynamics and analytically express the excited-state probability and von Neumann entropy for a single qubit and the concurrence for the two-qubit system. The system with qubit frequency lying inside PBG exhibits strong non-Markovian nature due to the formation of photon bound state with the qubit. This strong non-Markovian nature leads to the long-time memory effect of the PBG reservoir on the excited-state qubit. Systems with qubit frequency lying inside the PBG experience non-Markovian processes of operation while those outside the PBG will experience stochastic processes. The deeper inside the PBG the system is detuned, the higher degree of non-Markovianity the system exhibits. The long-time memory effect leading to this non-Markovianity leads to the preservation of entanglement between the single qubit and the PBG reservoir through the steady entropy with non-zero value. The large value of the steady entropy ensures that the quantum information can be stored in the qubit for a long period of time. This long-time memory effect disappears as the qubit frequency lying outside the PBG region.

As the two-qubit system is concerned, the results of concurrence dynamics reveal how the effect of PBG reservoir controls the entanglement through the initial prepared conditions and the degree of non-Markovianity. The entanglement mechanism for the prepared states $\left|\Phi\right\rangle$ is interchanging the trapped photon and while for the prepared $\left|\Psi\right\rangle$ state is dipole-dipole interaction. The PBG reservoir provides the qubits a better environment for preserving entanglement through interchanging the trapped photon than through dipole-dipole interaction. How the degree of non-Markovianity affects the entanglement is displayed in the steady values of concurrence with respect to the detuning frequency of the qubit. The deeper the qubit frequency is detuned into the PBG, the higher degree of the non-Markovianity and the stronger entanglement trapping yields. The system exhibits the non-Markovian nature of achieving entanglement only for the qubit frequency located inside the PBG. Unphysical entanglement trapping does not exist in the system with the qubit frequency lying outside the PBG region. These new predictions of the entropy and concurrence dynamics are different from those of the previous theoretical studies which predicted the existence of entanglement trapping in the system with qubit frequency lying outside the PBG region. The accuracy of these predictions is based on the appropriate mathematical method of fractional calculus for non-Markovian dynamics and reasonable results of physical phenomenon. The results about how the PBG reservoir affects the qubit's dynamics can be applied directly to the experimental systems of artificial crystals with PBG structured reservoir. We believe the entanglement mechanisms between the qubits and PBG reservoir hold the key technology of applying the PBG materials to the primitive blocks of quantum computers.

\begin{acknowledgments}
We would like to gratefully acknowledge partially financial support from the National Science Council (NSC), Taiwan under Contract Nos. NSC 100-2811-M-034-003, NSC 99-2112-M-006-017-MY3, NSC 99-2221-E-009-095-MY3, and NSC-99-2112-M-034-002-MY3.
\end{acknowledgments}
%\bibliographystyle{unsrt}

%\bibliography{2012QInAPC}

\appendix
\section{CALCULATION OF NON-MARKOVIAN DYNAMICS OF SINGLE QUBIT SYSTEM THROUGH FRACTIONAL CALCULUS}
The equation of motion of this qubit system in simplified form is given in Eq. (9). We use the fractional time derivative, one of the operators of fractional calculus, to express the integral term of this kinetic equation. In the well-known Riemann-Liouvile definition, the fractional time derivative operator $d^{\nu}/dt^{\nu}$ is expressed as \cite{IPodlubny99}
\begin{equation}
\frac{d^{\nu}}{dt^{\nu}}f(t)=\frac{1}{\Gamma (n-\nu)}\frac{d^{n}}{dt^{n}}\int^{t}_{a}\frac{f(\tau)}{(t-\tau)^{\nu-n+1}}d\tau
\end{equation}
for $n\leq\nu<n+1$ and $\Gamma (x)$ being the Gamma function. The integral term of Eq. (9) $\int^{t}_{0}\frac{U_{p}(\tau)}{(t-\tau)^{3/2}}d\tau$ can be expressed as this fractional operator with order $\nu=1/2$ and $n=0$. Through this expression and applying the integral operator ($d^{-1}/dt^{-1}$) and fractional differentiation operator $d^{1/2}/dt^{1/2}$ to Eq. (9), the fractional kinetic equation is obtained as  
\begin{equation}
\frac{d^{1/2}}{dt^{1/2}}U_{p}(t)+i\delta\frac{d^{-1/2}}{dt^{-1/2}} U_{p}(t)+\frac{2\beta^{1/2}e^{i\pi/4}}{f^{3/2}}U_{p}(t)=\frac{t^{-1/2}}{\sqrt{\pi}}.
\end{equation}
This kinetic equation governing the future of the qubit indicates a subordinated stochastic process directing to a stable probability distribution because of its interacting with the PBG reservoir. This fractional differential equation can be solved analytically through the fractional Laplace transform. The basic formula used here is
\begin{equation}
\textsl{L}\left\{\frac{d^{\nu}}{dt^{\nu}}f(t)\right\}\equiv\int^{\infty}_{0}e^{-st}\frac{d^{\nu}}{dt^{\nu}}f(t)dt=s^{\nu}\textsl{L}\left\{f(t)\right\}-\sum^{n-1}_{m=0}s^{m}\left[\frac{d^{\nu-m-1}}{dt^{\nu-m-1}}f(t)\right]_{t=0}
\end{equation}
with the Laplace variable $s$. By performing the fractional Laplace transform on Eq. (9), we can obtain
\begin{equation}
\tilde{U}_{p}(s)=\frac{1}{s+i\delta+2\beta^{1/2}e^{i\pi/4}s^{1/2}/f^{3/2}}
\end{equation}
where $\tilde{U}_{p}(s)$ is the Laplace transform of ${U}_{p}(t)$. As we express the denominator part of this Laplace transform as a sum of partial fractions through the roots of the indicial equation $Y^{2}+2\beta^{1/2}e^{i\pi/4}Y/f^{3/2}+i\delta=0$ with the variable $Y$ converted from $s^{1/2}$, the analytical solution of the Eq. (A2) can be obtained by means of the inverse Laplace transform. There exist two kinds of roots in the indicial equation: degenerate root and non-degenerate ones. For the degenerate case with $\beta/f^{3}=\delta$, the partial-fractional form of $\tilde{U_{p}}(s)=1/\left(\sqrt{s}+\frac{\beta^{1/2}e^{i\pi/4}}{f^{3/2}}\right)^{2}$ while that for the non-degenerate case with $\beta/f^{3}\neq\delta$ is $\tilde{U_{p}}(s)=\left[\frac{1}{(\sqrt{s}-Y_{1})}-\frac{1}{(\sqrt{s}-Y_{2})}\right]\frac{1}{(Y_{1}-Y_{2})}$ with $Y_{1}=e^{i\pi/4}\left(-\frac{\beta^{1/2}}{f^{3/2}}+\sqrt{\frac{\beta}{f^{3}}-\delta}\right)$ and $Y_{2}=e^{i\pi/4}\left(-\frac{\beta^{1/2}}{f^{3/2}}-\sqrt{\frac{\beta}{f^{3}}-\delta}\right)$. The inverse Laplace transform of these partial fractions with fractional power of $s$ can be found only in the books of fractional calculus which give $\textsl{L}^{-1}\left\{\frac{1}{(s^{\nu}-a)}\right\}=\sum^{q}_{j=1}a^{j-1}E_{t}(j\nu-1,a^{q})$ and  $\textsl{L}^{-1}\left\{\frac{1}{(s^{\nu}-a)^{2}}\right\}=\sum^{q}_{j=1}\sum^{q}_{m=1}a^{j+m-2}\left\{tE_{t}\left((j+m)\nu-2,a^{q}\right)-\left[(j+m)\nu-2\right]E_{t}\left((j+m)\nu-1,a^{q}\right)\right\}$ with $\nu=1,1/2,1/3,...$ and $q=1,2,...,1/\nu$. Here the fractional exponential function $E_{t}(\alpha,a)$ is defined as the fractional derivative of an ordinary exponential function $E_{t}(\alpha,a)\equiv\frac{d^{-\alpha}}{dt^{-\alpha}}e^{at}=t^{\alpha}\sum^{\infty}_{n=0}\frac{(at)^{n}}{\Gamma (\alpha+n+1)}$ with the derivative formula $\frac{d^{\mu}}{dt^{\mu}}E_{t}(\alpha,a)=E_{t}(\alpha-\mu,a)$. A linear combination of the fractional exponential functions is a potential solution of the fractional differential equation because its functional form will not be changed after being performed derivative operator with fractional or integral order. By applying the inverse Laplace transforms to the partial fractions of $\tilde{U}_{p}(s)$, we can obtain the analytical solution of Eq. (9) as
\begin{equation}
U_{p}(t)=\frac{1}{2e^{i\pi/4}\sqrt{\beta/f^{3}-\delta}}\times\left[Y^{2}_{1}E_{t}(1/2,Y^{2}_{1})-Y^{2}_{2}E_{t}(1/2,Y^{2}_{2})+Y_{1}e^{Y^{2}_{1}t}-Y_{2}e^{Y^{2}_{2}t}\right],
\end{equation}
for the non-degenerate case with $\beta/f^{3}\neq\delta$; and
\begin{equation}
U_{p}(t)=-2\frac{\beta^{3/2}e^{i3\pi/4}}{f^{9/2}}tE_{t}(\frac{1}{2},i\beta/f^{3})-\frac{\beta^{1/2}e^{i\pi/4}}{f^{3/2}}E_{t}(\frac{1}{2},i\beta/f^{3})
+(1+\frac{2it\beta}{f^{3}})e^{i\beta t/f^{3}}-2\frac{\beta^{1/2}e^{i\pi/4}}{f^{3/2}\sqrt{\pi}}t^{1/2}
\end{equation}
for the degenerate case with $\beta/f^{3}=\delta$ which have been articulated in Eqs. (10) and (11) of the text.

%%%%%%%%%%%%%%%%%%%%%%%%%%%%%%%%%%%%%%%%%%%%%%%%%%%%%%%%%%%%%%%%%%%%%%%%%%%
\begin{figure}[hbt]

% \begin{tabular}{c c}
%\includegraphics[bb=0 0 757 357, width=20.0cm, clip]{1.png}
\subfigure[]{	\includegraphics[width=5.0cm, bb=0 0 553 744, clip]{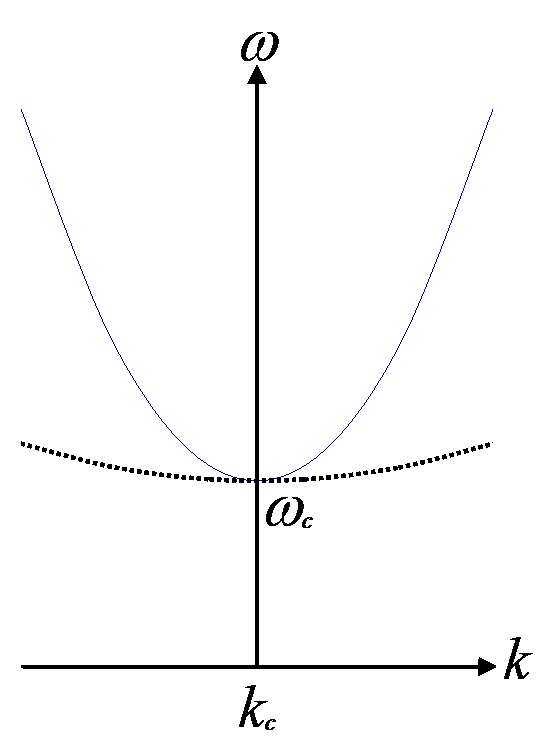}}  % &
\subfigure[]{	\includegraphics[width=11.0cm, bb=0 0 734 438, clip]{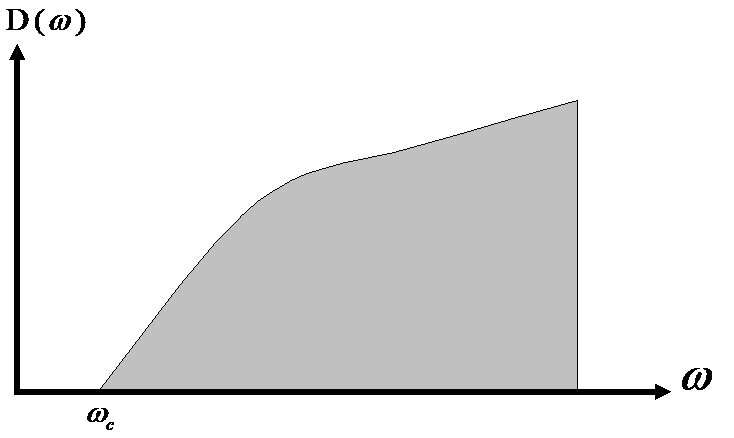}} %\\
%\end{tabular}
\caption{(Color online) (a) Dispersion relation near the band edge of a PBG material with directional dependent curvature is expressed by the effective-mass approximation with the edge frequency $\omega_{c}$. (b) Photon DOS $D(\omega)$ of the anisotropic PBG reservoir exhibiting cut-off photon mode below the edge frequency $\omega_{c}$. }
\end{figure}
%%%%%%%%%%%%%%%%%%%%%%%%%%%%%%%%%%%%%%%%%%%%%%%%%%%%%%%%%%%%%%%%%%%%%%%%%%%%%%%%%%

%%%%%%%%%%%%%%%%%%%%%%%%%%%%%%%%%%%%%%%%%%%%%%%%%%%%%%%%%%%%%%%%%%%%%%%%%%%%%%%%%%%%%%%

\begin{figure}
\subfigure[]{	\includegraphics[viewport=0 0 1507 2048 , width=0.48\textwidth,  bb=0 0 2048 1507]{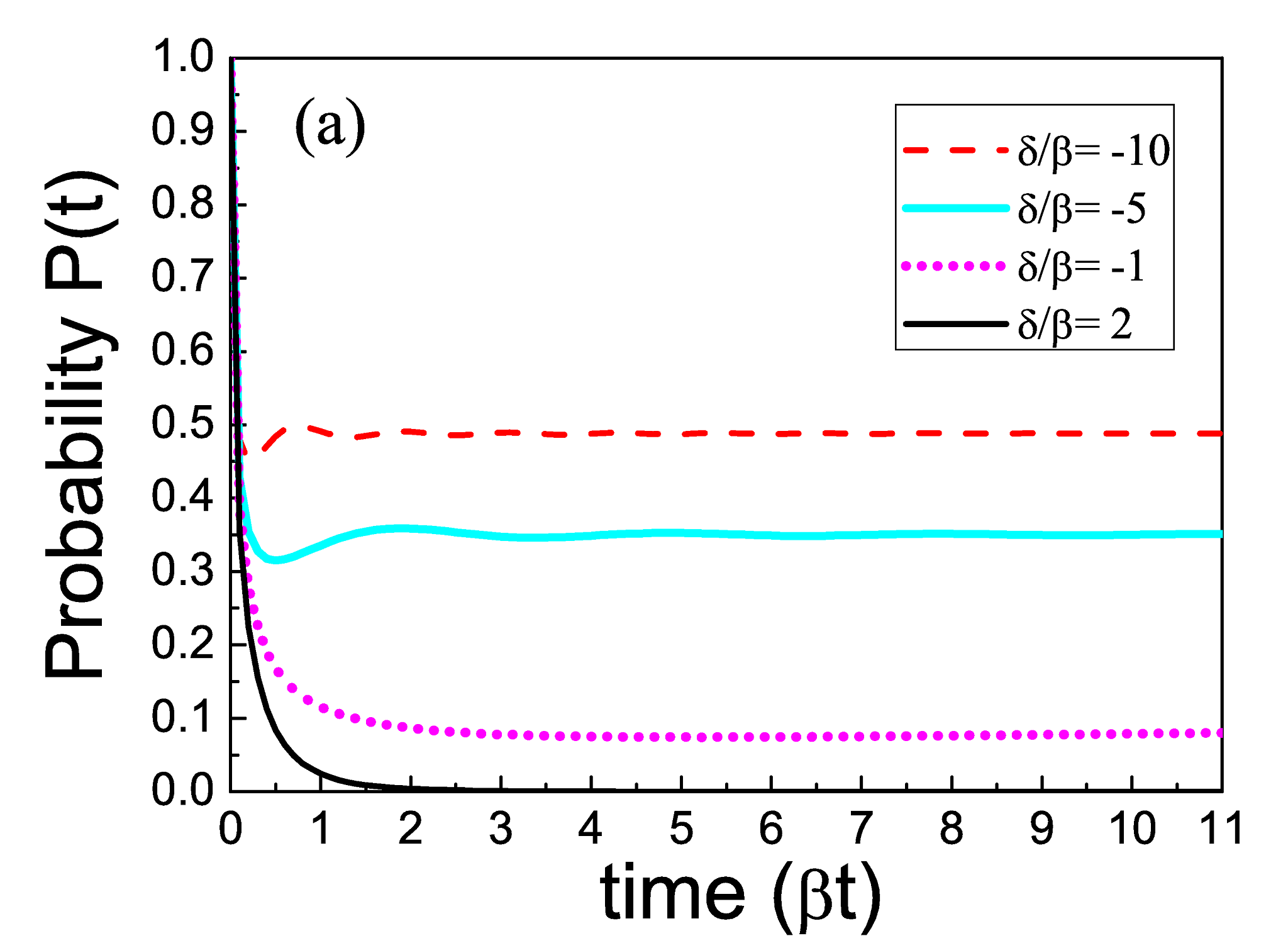}}
\subfigure[]{	\includegraphics[viewport=0 0 1519 2048 , width=0.48\textwidth,  bb=0 0 2048 1519]{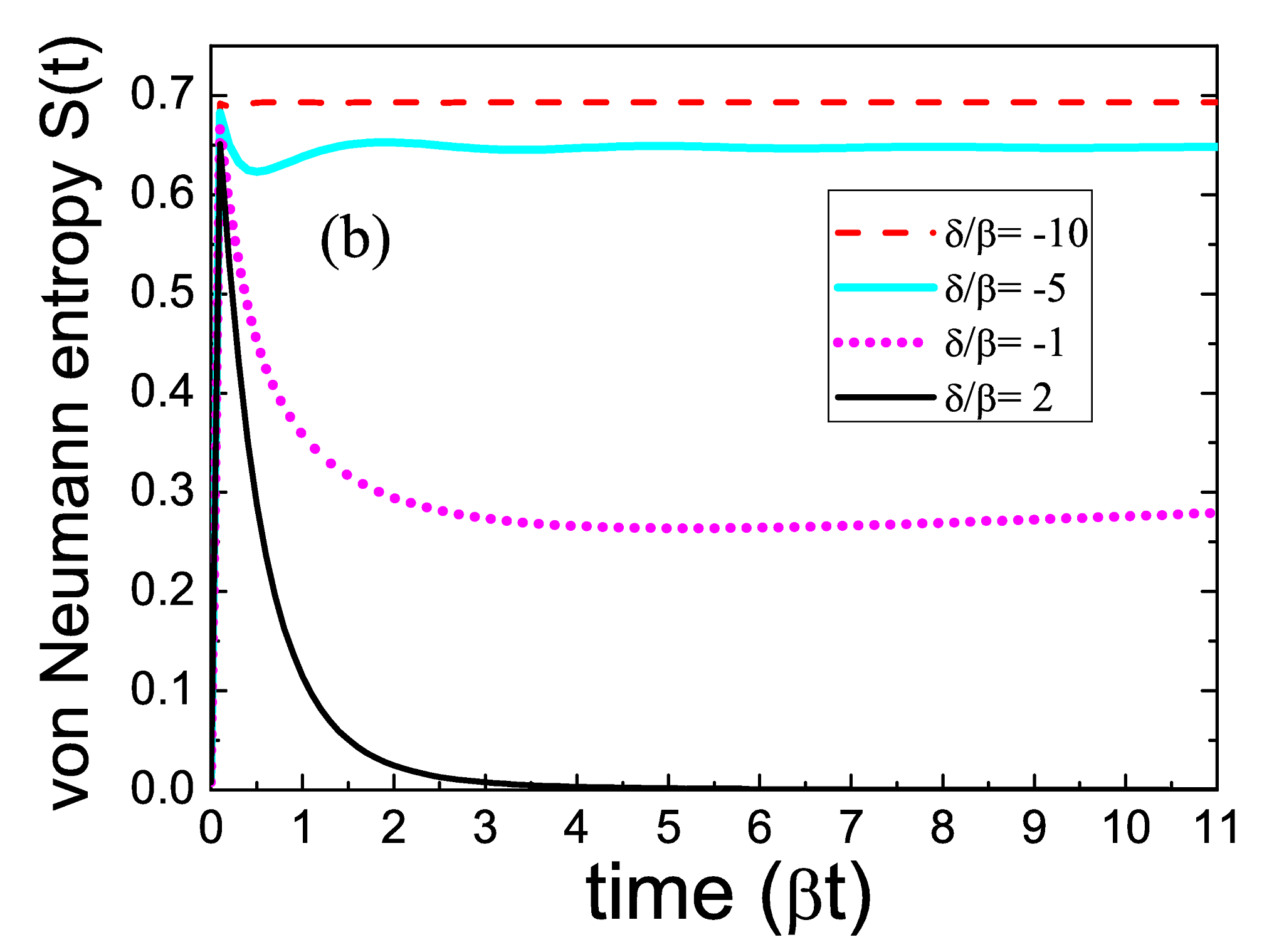}}
\caption{ (Color online)  Dynamics of single qubit in PBG reservoir. (a) Qubit's excited-state probability $P(t)=\left|U_{p}(t)\right|^{2}$ with different detuning frequencies $\delta/\beta=(\omega_{10}-\omega_{c})/\beta$ from the band edge frequency $\omega_{c}$. (b)von Neumann entropy $S(t)=-Tr\left[\hat{\rho}^{s}(t) log\hat{\rho}^{s}(t)\right]=-\lambda_{+}log\lambda_{+}-\lambda_{-}log\lambda_{-}$ of the non-Markovian ($\delta/\beta<0$) and Markovian ($\delta/\beta=2$) systems. }
\end{figure}

%%%%%%%%%%%%%%%%%%%%%%%%%%%%%%%%%%%%%%%%%%%%%%%%%%%%%%%%%%%%%%%%%%%%%%%%%%%%%%%%%%%%%

%%%%%%%%%%%%%%%%%%%%%%%%%%%%%%%%%%%%%%%%%%%%%%%%%%%%%%%%%%%%%%%%%%%%%%%%%%%%%%%%%%%%%%%

\begin{figure}

\subfigure[]{	\includegraphics[bb=0 0 606 483, width=8.0cm,  clip]{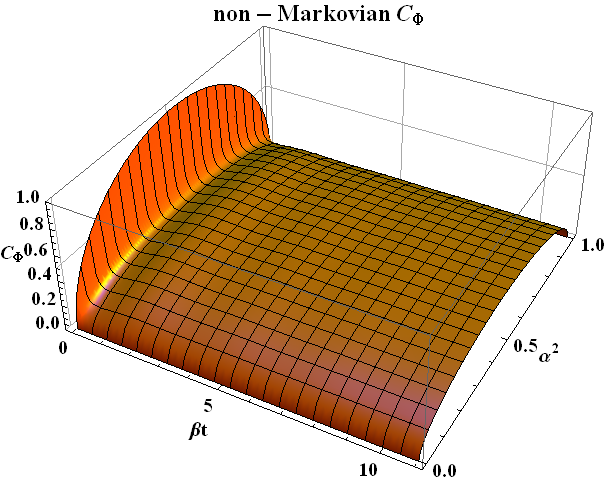}}
\subfigure[]{	\includegraphics[bb=0 0 600 479, width=8.0cm,  clip]{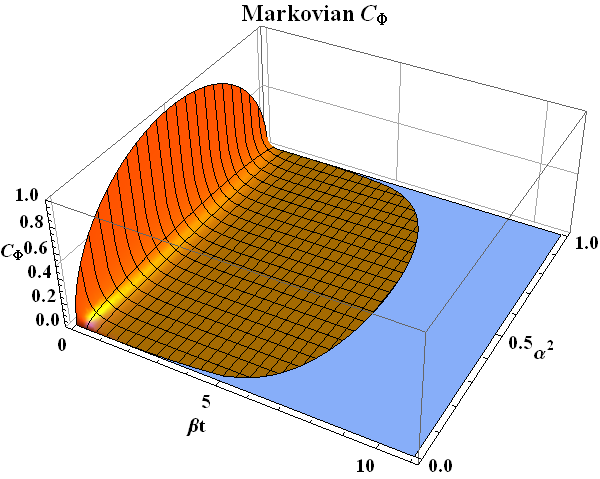}}
\subfigure[]{	\includegraphics[bb=0 0 601 480, width=8.0cm,  clip]{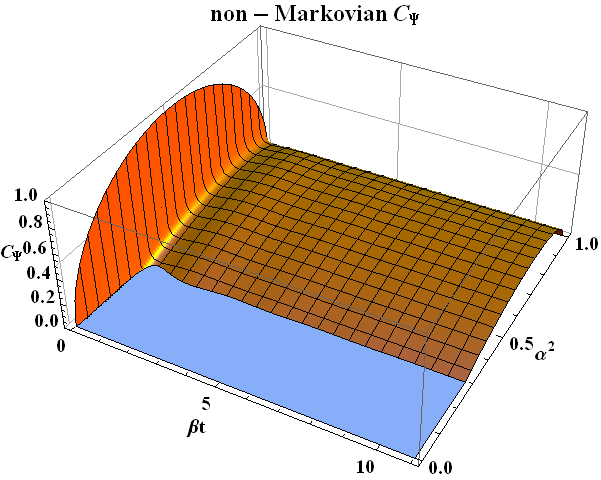}}
\subfigure[]{	\includegraphics[bb=0 0 602 481, width=8.0cm,  clip]{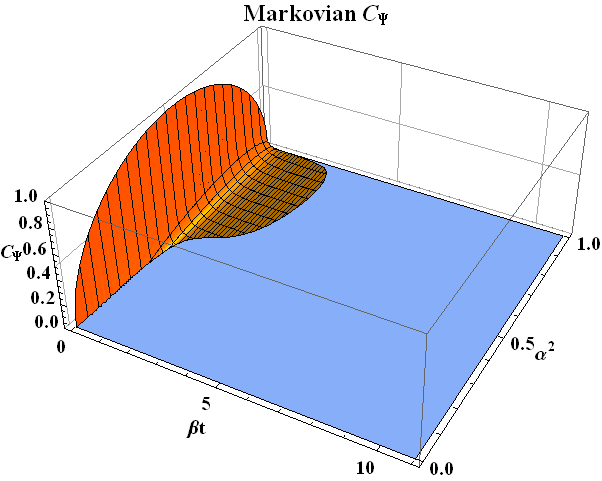}}

\caption{ (Color online)  Variation of concurrence dynamics with respect to the initial entangled degree $\alpha$ for the prepared states $\left|\Phi\right\rangle=\alpha\left|01\right\rangle+\sqrt{1-\alpha^{2}}\left|10\right\rangle$ and $\left|\Psi\right\rangle=\alpha\left|00\right\rangle+\sqrt{1-\alpha^{2}}\left|11\right\rangle$. (a) Concurrence for the prepared $\left|\Phi\right\rangle$ states with qubit frequency lying inside the PBG (non-Markovian system, $\delta/\beta=-5$) and (b) outside PBG (Markovian system, $\delta/\beta=2$);  (c) Concurrence for the prepared $\left|\Psi\right\rangle$ states with qubit frequency lying inside the PBG (non-Markovian system, $\delta/\beta=-5$) and (d) outside PBG (Markovian system, $\delta/\beta=2$)}
\end{figure}

%%%%%%%%%%%%%%%%%%%%%%%%%%%%%%%%%%%%%%%%%%%%%%%%%%%%%%%%%%%%%%%%%%%%%%%%%%%%%%%%%%%%%

%%%%%%%%%%%%%%%%%%%%%%%%%%%%%%%%%%%%%%%%%%%%%%%%%%%%%%%%%%%%%%%%%%%%%%%%%%%%%%%%%%%%%%%

\begin{figure}
\subfigure[]{	\includegraphics[viewport=0 0 1492 2048 ,bb=0 0 2048 1492, width=0.48\textwidth,  clip]{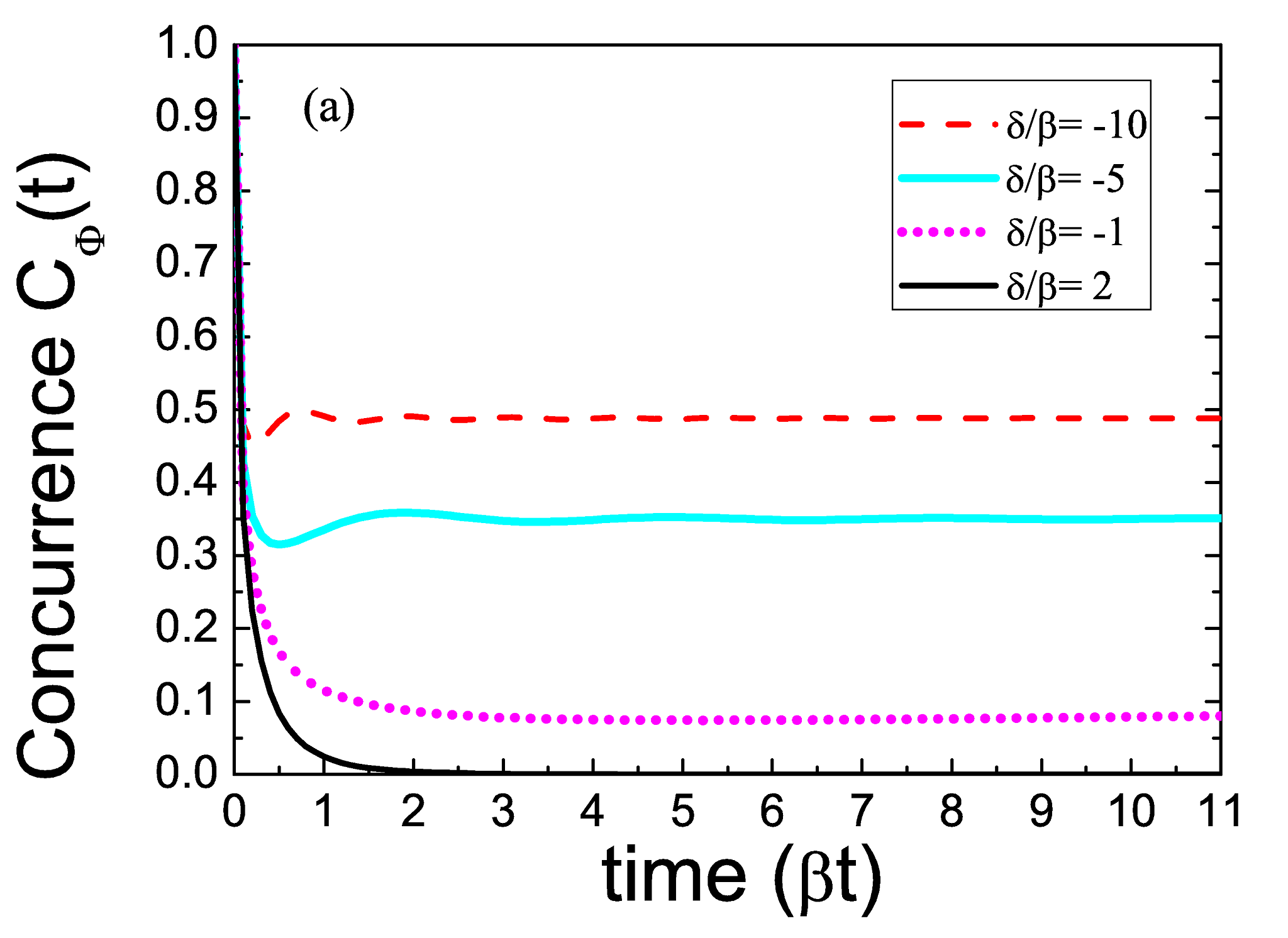}}
\subfigure[]{	\includegraphics[viewport=0 0 1551 2048 ,bb=0 0 2048 1551, width=0.48\textwidth,  clip]{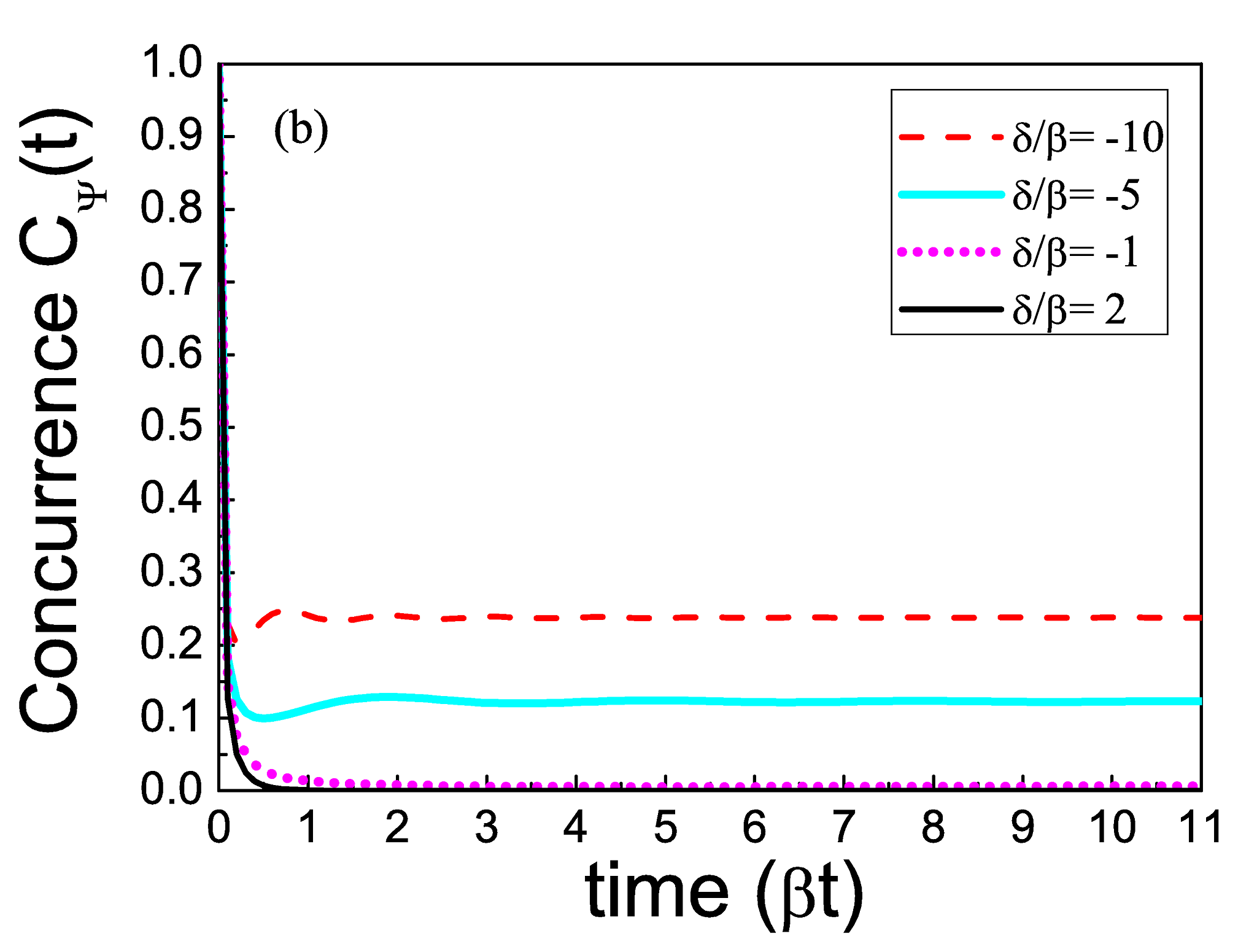}}

\caption{ (Color online) Concurrence dynamics of the two-qubit system for the prepared Bell states $\left|\Phi\right\rangle_{B}=(\left|01\right\rangle\pm\left|10\right\rangle)/\sqrt{2}$ in (a) and for $\left|\Psi\right\rangle_{B}=(\left|11\right\rangle\pm\left|00\right\rangle)/\sqrt{2}$ in (b). Steady values of concurrence in (a) and (b) with respect to the qubit detuning frequency $\delta/\beta$}
\end{figure}

%%%%%%%%%%%%%%%%%%%%%%%%%%%%%%%%%%%%%%%%%%%%%%%%%%%%%%%%%%%%%%%%%%%%%%%%%%%%%%%%%%%%%

\end{document}